# Thermo-optically Reconfigurable PbTe Mie Resonator Meta-atoms

Tomer Lewi[1], Hayden A. Evans[2], Nikita A. Butakov[1] and Jon A. Schuller[1*]

[1] Department of Electrical and Computer Engineering, University of California, Santa Barbara, California 93106, USA
[2] Materials Research Laboratory and Department of Chemistry and Biochemistry, University of California, Santa Barbara, California 93106, USA

**Subwavelength Mie resonators have enabled new classes of optical antennas[1-4], photodetectors[5,6], antireflection coatings[7], and magnetic mirrors[8] and can act as the basic meta-atoms constituents of low-loss dielectric metasurfaces[2-4,8-10]. In any application, tunable Mie resonances are key to achieving dynamic and reconfigurable operation[11,12]. Sub-linewidth tuning has been achieved via coupling to liquid crystals[13], ultrafast free-carrier injection[14,15,16,17], stretchable substrates[18] and with phase change materials[19]. Here, we demonstrate ultra-wide dynamic tuning of PbTe Mie resonators fabricated via both laser ablation and a novel solution-processing approach. Taking advantage of the extremely large thermo-optic (TO) coefficient and high refractive index of PbTe, we demonstrate high-quality factor Mie-resonances that are tuned by several linewidths with temperature modulation as small as ΔT~10K. When combined into metasurface arrays these effects can be exploited in ultra-narrow active notch filers and metasurface phase shifters that require only few-kelvin modulation. These findings demonstrate the enabling potential of thermo-optically tunable PbTe meta-atoms and metasurfaces.**

Semiconductor Mie resonators and metasurface constituents have shown great promise as enabling components of nanophotonic technologies. Absorbing resonators can directly generate electronic response in e.g. detectors[5,6] and solar cells[20,21]. In non-absorbing materials, high quality factor (Q) Mie resonances can strongly enhance nonlinear processes[22,23] while ultra-low losses enable high radiation efficiency antennas and transmissive metasurfaces[24,25]. An example of a Mie resonator is shown in Figure 1a. A spherical Ge nanoparticle, fabricated via femtosecond laser ablation[11,26,27] (see methods), exhibits a sequence of Infrared scattering resonances in both analytical calculations (red) and single particle FTIR reflection spectra (black). In principle, resonances can be tuned via the thermo-optic effect (TOE), i.e. refractive index variation with temperature dn/dT. However, typical applications of TOE exploit small index changes acting over distances much larger than a wavelength to achieve useful modulation. The moderate Qs and subwavelength size of semiconductor Mie resonators and meta-atoms necessitates extreme TO tunability. For instance, consider the fundamental magnetic dipole (MD) mode in Figure 1a. The MD Mie resonance wavelength is defined by an eigenvalue equation—$2\pi rn/\lambda=3.077$ (r is the particle radius, n is the refractive index and λ is the free space wavelength)—and exhibits a moderate $Q_{Ge}$ ~ 7.5. To shift the resonance by one linewidth requires an index shift of ≈0.6.

In Figure 1(b) we plot the room temperature (RT) TO coefficient of Ge (blue) and PbTe (black) (see supplementary section 1)[28]. Ge has the highest reported TO coefficient of any semiconductor outside the lead chalcogenide family and is a natural choice for investigating TO tunability (the TO coefficient of Si is plotted for comparison)[28]. Tuning the Ge MD resonance of Figure 1(a) by one linewidth requires an extraordinary temperature swing of ΔT=1500K. For practical and efficient modulation, materials with far higher TO coefficients are needed. The lead chalcogenide family of group IV-VI semiconductors, have extremely large TO coefficients. Amongst them, PbTe possesses the highest amplitude. Its room temperature dn/dT value, plotted in in Figure 1(b), is ~5 times larger than Ge. Interestingly, the sign of the TO coefficient is negative (dn/dT<0) due to anomalous temperature-dependent bandgap dispersion[28-30] (supplementary section 1). This large TO coefficient makes PbTe an intriguing candidate material for TO-tunable Mie resonators and metasurfaces.

Experimental (black) and calculated (red) scattering spectra of a PbTe Mie resonator (inset) are shown in Figure 1(c). In comparison with Ge, two important characteristics can be observed, both arising from the larger refractive index ($n_{PbTe}$~6, $n_{Ge}$~4). First, although the PbTe sphere size is smaller, resonances occur at longer wavelengths with higher calculated scattering efficiencies. Secondly, all PbTe resonances exhibit higher Qs than their Ge counterparts. For instance the PbTe MD mode ($Q_{PbTe}$=10.8) is almost 50% narrower than in Ge ($Q_{Ge}$=7.5). The anomalously large refractive index and TO coefficient of PbTe enable the TO-tunable nanophotonic elements described here. Experimental measurements of temperature-dependent Ge and PbTe Mie resonances are shown in Figure 1d. In going from 80K (red) to 293K (orange) the Ge MD mode red-shifts by 90nm. Dividing by the linewidth (FWHM), this corresponds to a normalized tuning of 0.09. In comparison, the same temperature swing causes a significantly larger *blue-shift* (480nm), and normalized tuning (-0.65) in PbTe.  These results clearly demonstrate the strong, negative TO effect in PbTe nanoparticles and is the largest dynamic tuning of Mie resonators reported to date.[13-16] Building from these results, we demonstrate a further ~ 100-fold increase of normalized tunability in subwavelength laser-fabricated and solution-processed PbTe resonators. We show that these affects arise from a combination of increased TO coefficient at low temperatures and ultra-narrow high-order Mie resonances. Finally, we conclude with demonstrations of TO tunable filters and metasurfaces enabled by these novel phenomena.

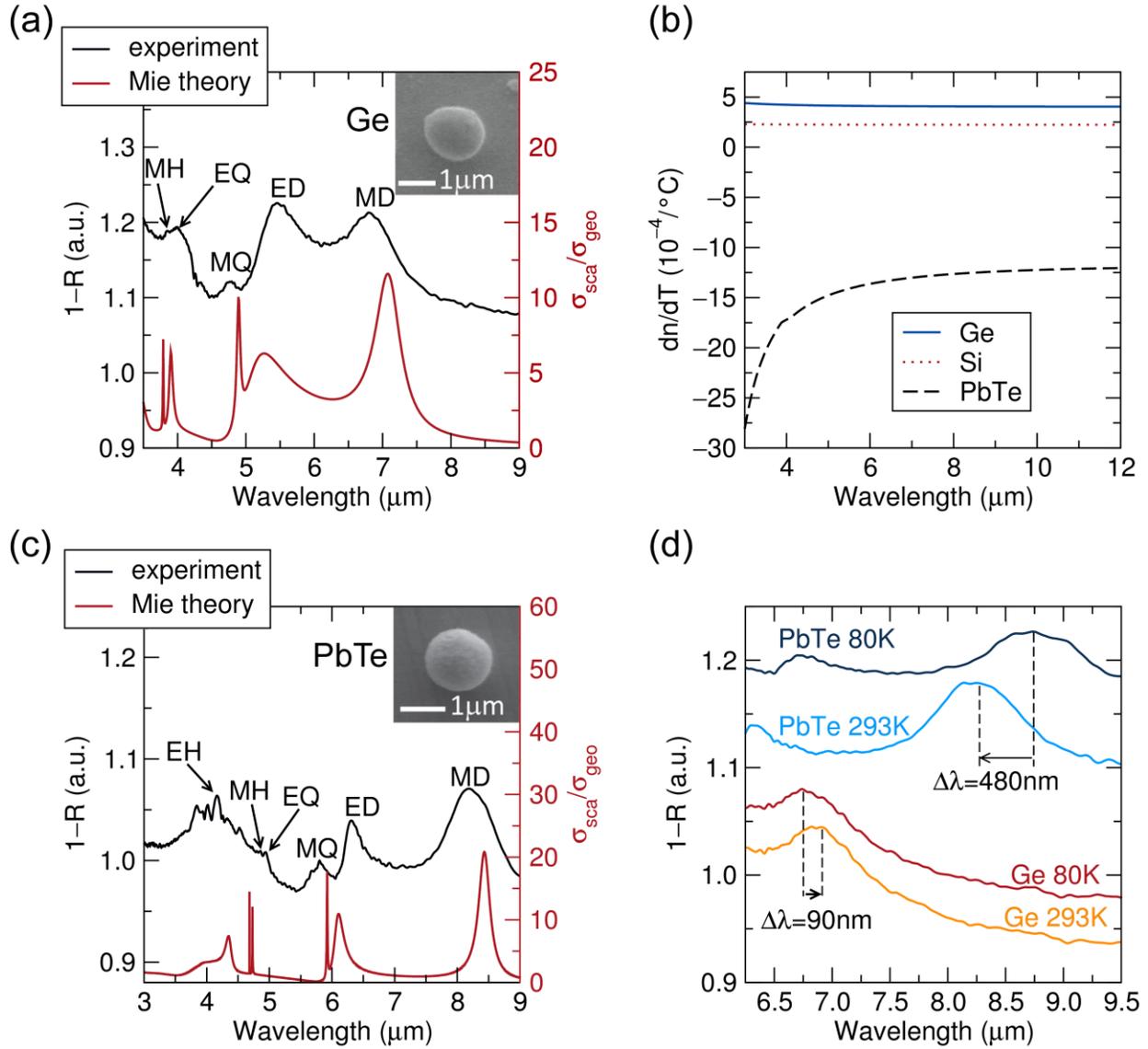

Figure 1: Large thermo-optic tuning in Ge and PbTe single resonators: (a) & (c) Infrared spectra of a d=1.7µm Ge sphere on Ge substrate (a) and a d=1.45µm PbTe sphere on PbTe substrate (c), respectively. Experimental spectra show good agreement with the calculated Mie scattering cross-sections $\sigma_{sca}$ normalized to the geometric cross-section $\sigma_{geo}$. The multipolar resonances are identified and labeled as magnetic and electric dipoles (MD, ED), quadrupoles (MQ, EQ) and hexapoles (MH, EH). The insets shows SEM images of the particles. (b) TO coefficient of Ge, Si and PbTe at RT. (d) TO induced shift between 80K and 293K in the MD mode of the d=1.45µm PbTe sphere compared to the d=1.7µm Ge sphere. The combination of a higher Q and larger TO coefficient enables far greater tunability in PbTe.

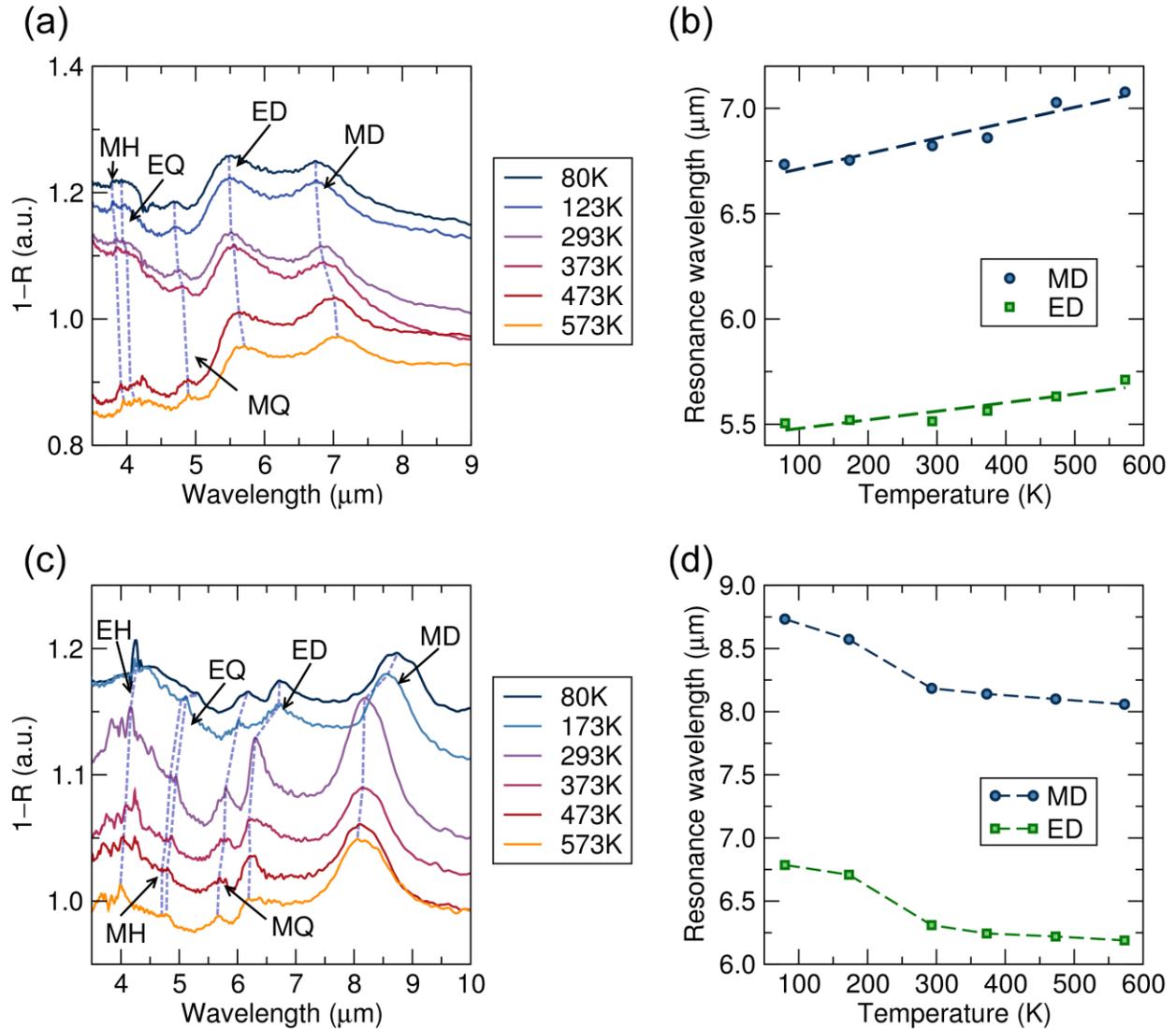

Figure 2: Multipolar resonance tuning and induced index shift in Ge and PbTe nanospheres. (a)&(c) Temperature dependent multipolar resonance shifts in Ge (a) and PbTe (c) spheres of diameters d=1.7μm and 1.45μm respectively. (b)&(d) MD and ED resonance wavelength dispersion with temperature for Ge (b) and PbTe (d). The dashed lines in (b) are linear fits to the data, whereas the dashed lines in (d) are a guide to the eye. Compared to Ge, PbTe exhibits a larger TO coefficient, with opposite sign, and a marked increase in magnitude below room temperature.

Tracking the shift of all Mie resonances across a variety of temperatures provides valuable insight into the dispersion of TOE with temperature and wavelength. Ge modes up to hexapole order are labeled in Figure 2a for a series of temperatures between 80-573 K. The MD (blue) and ED (green) resonance wavelengths are plotted in Figure 2(b). The modes exhibit a uniform linear shift with temperature, in excellent agreement with Mie theory calculations (supplementary

Figure S3)[11] incorporating literature values of the TO coefficient (Figure 1(b)). The higher order modes follow similar behavior as the TOE and refractive index of Ge are nearly constant across the MIR (supplementary Figure S3). However, higher-order modes have larger Qs and can be tuned by more than a linewidth with temperature swings as low as ~120K (supplementary Figures S3 and S5). The tunable PbTe resonances (Figure 2(c) and Figure 2(d)), exhibit different characteristics. The chromatic dispersion of PbTe's TO coefficient (Figure 1a) is observed in these measurements (supplementary Figure S4). Induced index shifts as large as Δn≈0.7 were obtained at shorter wavelengths and are expected to increase for wavelengths approaching the excitonic bandgap (supplementary Figure S4)[28]. Most importantly, the wavelength shifts at low temperature (80K-293K) are much larger than the shifts above room temperature (293K-573K), as evidenced by the differences in slope seen in Figure 2d. These results show that the largest dn/dT occurs between 80K to RT and a more thorough examination reveals a maximum in dn/dT somewhere between 80K and 173K (supplementary Figure S3). Although observed shifts are consistent with other measurements,[29,31,32] this significant increase in TO coefficient at low temperatures has not previously been reported. Standard TO models[28], based on temperature-dependent bandgap ($E_g$) dispersion,[30,33] are unable to explain the magnitude of this effect (supplementary Figures S3 and S4), suggesting the possibility that unknown physical mechanisms are at play. Regardless, by operating at cryogenic temperatures, significant increases in TO tunability can be achieved.

Combining the large low-temperature TOE with high-Q resonances enables complete switching (i.e. tuning by more than one linewidth) of resonances with significantly reduced temperature swings (ΔT). Due to its high index and negligible losses, ultrahigh-Q resonances are observed in single PbTe resonators. TO spectral tuning of a d=1.76μm PbTe sphere is shown in Figure 3a (SEM image in Figure 3b). As expected, all the resonances are blue shifted with increasing temperature. Unlike Figure 1, however, here we focus on the pronounced high-Q modes. Most notably, a very sharp dip with Q ~100 is observed at ~λ=4.8μm and is marked by the shaded orange area. This sharp resonance dip is shifted by more than 8 linewidths over the 80-573K temperature range. Calculated RT electric and magnetic field (Figure 3(b)) reveal the combined contribution of electric hexapole (EH) and ED resonances to this spectral feature (also supplementary Figure S6). The experimental dip position (black) shows excellent agreement with FDTD simulations (red solid) and Mie theory (red dashed) (Figure 3c). As seen in Figure 3d, this sharp resonance can be tuned by more than one linewidth (normalized tunability=1.6) with a temperature swing as small as ΔT=10K. This exceptional tunability is enabled by the combination of the narrow linewidth and the enhanced TOE. The observed dn/dT at 173K (-0.0134 $K^{-1}$) is almost an order of magnitude larger than previously reported RT TO effects[28] (dn/dT=-0.0015$K^{-1}$).

Implementations of these tunable effects in devices and metasurfaces will ultimately require easy and scalable fabrication methods. For instance, typical Mie resonator based metasurfaces and devices are fabricated with conventional semiconductor processing techniques[2-4,7-10]. The use of PbTe resonators, on the other hand, provides an intriguing alternative: solution-processing. For instance, solution processed PbTe photovoltaics[34] and thermoelectrics[35] have been demonstrated. Using solution processing (methods) we synthesize cube-shaped PbTe resonators with sizes ranging between 0.4-3µm. An SEM (inset) and infrared spectra of an example resonator are shown in Figure 4a. When cast onto a high index semiconductor (Si) surface, the resonator exhibits a set of Mie resonances in experiments (black) and simulations (red) analogous to that in spherical particles. For instance, simulated field profiles (inset) confirm that the fundamental resonance at λ=7.13µm is an MD-like mode. Similarly, the extracted linewidth (Q~10) of the MD mode is comparable to that of spherical resonators. Dispersing such resonators onto metal substrates provides a convenient means to achieve even larger quality factors and improve the normalized tunabilty[36,37]. Spectra of a 1.78µm PbTe cube cast on a gold surface are shown in Figure 4(b). The interaction with mirror images[37] shifts all resonances to longer wavelengths. For instance, the usually circular symmetric magnetic field profile of the MD mode is truncated due to reflecting boundary conditions at the resonator-substrate interface (inset). Numerous high-Q (Q ~60) higher-order modes can also be seen—high Q features are not contingent on spherical geometries. Similar to spherical resonators (Figure. 3), these features can be tuned by several linewidths with a peak TO coefficient observed between 80-293 K (supplementary Figure S7 and S8). In fact, the observed index shifts are up to 15% larger than that in spherical resonators, perhaps due to a greater degree of crystallinity. Mie resonances for PbTe cubes of varying sizes are shown in Figure 4c. Features exhibit the expected linear size dispersion, and high quality resonances can be engineered across the infrared transparency range.

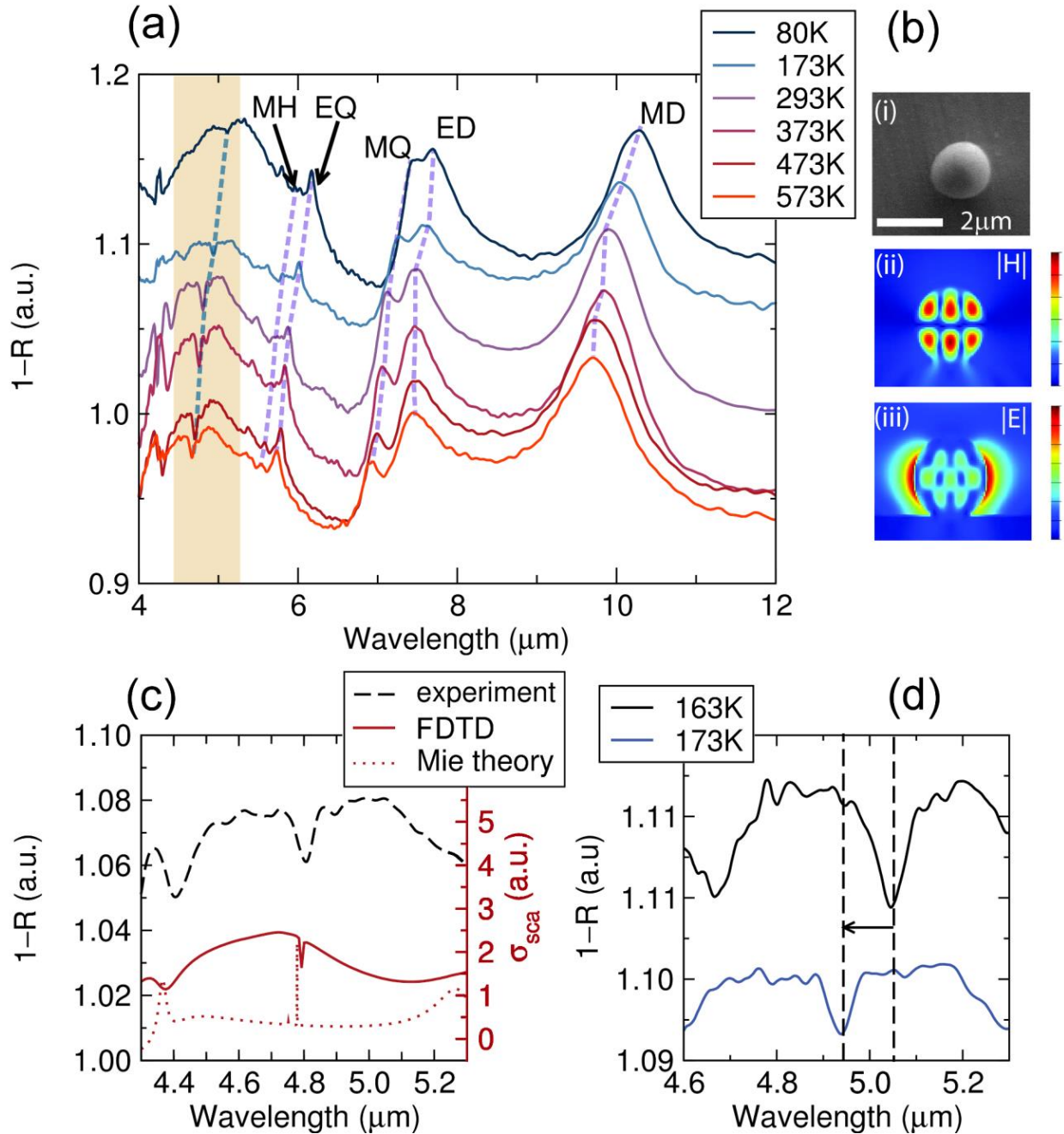

Figure 3: TO tuning of high quality factor (Q) resonances in PbTe particles. (a) Temperature dependent infrared spectra of a d=1.76μm PbTe sphere on PbTe substrate. The dashed lines are a guide to the eye. The light orange area marks the region of the sharp resonance dip and its corresponding tuning by multiple linewidths. (b) SEM image of the sphere along with magnetic and electric field profiles of the resonance dip at λ=4.79μm at RT. (c) Zoom-in on the RT high-Q resonance dip marked by the orange area in (a) along with Mie theory and FDTD simulation, showing excellent match between experiment and theory (d) Experimental demonstration of TO tuning by more than a linewidth, of the high-Q resonance dip with a temperature modulation (ΔT) of only 10K.

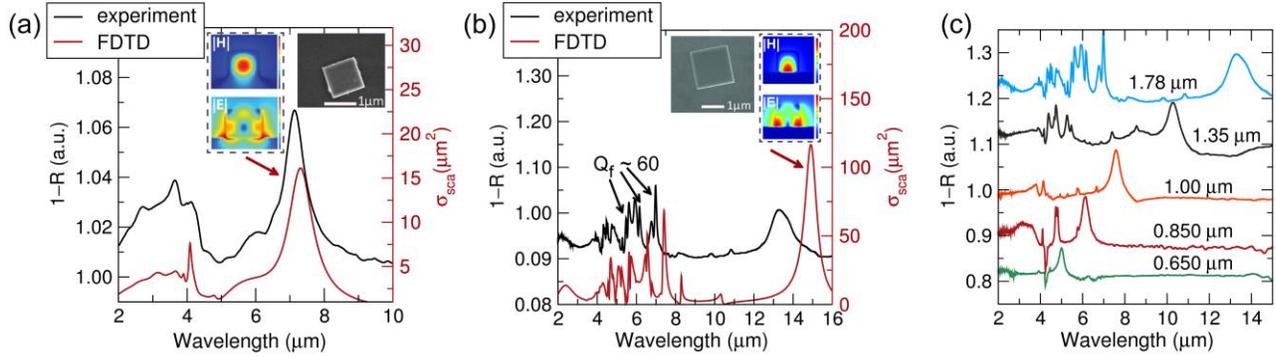

Figure 4: Tunable chemically synthesized PbTe cubic resonators: (a)&(b) Measured spectra and FDTD simulation of (a) 1 μm PbTe cube on Si substrate and (b) 1.78μm PbTe cube on gold substrate. The insets present SEM images of the cubes and the electric and magnetic field plots of the fundamental mode. (c) Geometric dispersion of PbTe cubes on gold demonstrating resonance scaling with size. Temperature dependent spectra are shown in the supplementary info (Figure S7 and S8).

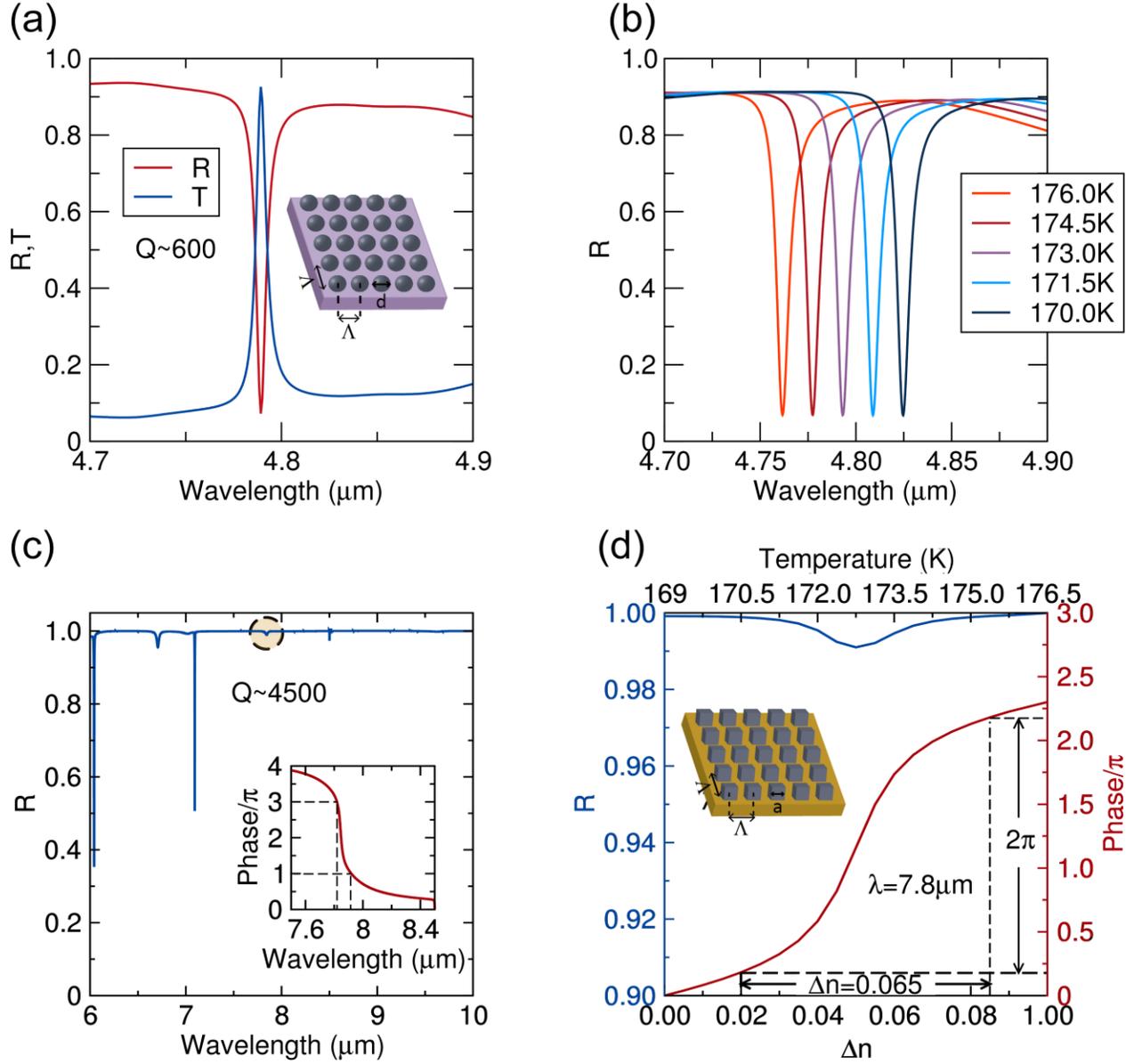

Figure 5: Ultra-high tunabilty in subwavelength PbTe metasurfaces. (a) Reflection and transmission FDTD simulation of a cubic spherical metasurface array (sphere diameter d=1.8μm and periodicity Λ=2μm) show a high-Q feature arising from the single particle mode discussed in Figure 3. (b) TO tuning of the reflection dip shown in (a). The dip is shifted by more than a linewidth (normalized tunability=1.975) with Δn=0.02—steps that corresponds to ΔT=1.5K intervals around T=173K. (c) Cubic metasurface arrays comprising PbTe cubes of side a=1.78μm on (PEC) substrate with periodicity of Λ=2μm exhibit multiple high-Q resonances. The inset shows how the phase evolves rapidly in the vicinity of a resonance at λ=7.8μm (marked by a black circle in the Reflection plot) (d) Reflection and phase at λ=7.8μm vs induced index Δn and temperature in the same cubic PbTe array described in (c). Up to 2π phase shift is achieved with greater than 99% reflectivity and Δn≤0.065 (corresponding to a temperature swing of ΔT~5K around T=173K).

The potential to exploit widely tunable PbTe meta-atoms in reconfigurable metasurfaces is illustrated in Figure 5. Simulated reflection and transmission spectra of a subwavelength 2D array of spheres (diameters of d=1.8μm and periodicity Λ=2μm) on a PbTe substrate is shown in Figure 5a. Reflection (transmission) dips (peaks) of ~90% can be observed due to coupling to the single-particle high-Q mode shown in Figure 3. Additionally, inter-particle coupling leads to resonance linewidth narrowing and much higher Qs are obtained (Q ~600). These high Q resonances now allow more than a linewidth of tunability (normalized tunability of 1.975) with an induced index shift of Δn=0.02 (Figure 5(b)). This induced index change corresponds to a temperature span of just ΔT=1.5K around T=173K, or equivalently ΔT=12K around RT. Similarly, a PbTe cube metasurface reflectarray (a=1.78 μm, Λ=2μm) on a perfect electric conductor (PEC) substrate exhibits sharp resonances with Qs as high as 4500 (Figure 5(c)) and linewidth tuning with Δn=0.02 (supplementary Figure S9). Moreover, utilizing the large associated phase shift around a resonance (circled feature), drastic phase modulation can be achieved. Figure 5(d) shows that 2π phase shift is obtained for Δn=0.065 (ΔT~5K around T=173K) while maintaining greater than 99% reflectivity. This ability to dynamically control reflection amplitudes and/or phase with small temperature modulation may enable a plethora of TO reconfigurable metasurface functionalities and devices.

In summary, we demonstrate ultra-wide dynamic tuning of infrared PbTe Mie resonator meta-atoms. Taking advantage of the extremely large TO coefficient and high refractive index of PbTe, we demonstrate high-Q Mie-resonances that are tuned by several linewidths with 10s of K temperature shifts. A detailed analysis of TO shifts from 80-573K reveal an approximately 10-fold increase in the TO coefficient below room temperature. This effect has not previously been reported and suggests the existence of additional TO tuning mechanisms. Demonstrations of similar high Q resonances and tunability in solution processed cubic resonators suggest new possibilities for meta-atom paints, coatings, etc. Simulations of particle arrays illustrate how these phenomena can be used to construct reconfigurable spectral filters and metasurfaces. These results highlight a path to new classes of TO tunable nanophotonic components and devices based on the unique properties of PbTe and suggest the importance of further investigations of tunable infrared properties in other IV-VI semiconductors.

Acknowledgments:

This work was supported by the Air Force Office of Scientific Research (FA9550-16-1-0393) and by the UC Office of the President Multi-campus Research Programs and Initiatives (MR-15-328528). Microscopy was performed with support from MRSEC Program of the NSF under Award No. DMR 1121053; a member of the NSF-funded Materials Research Facilities Network. Numerical calculations were performed with the support from the Centre for Scientific Computing from the CNSI, MRL: an NSF MRSEC (DMR-1121053), NSF CNS-0960316. N.A.B. acknowledges support from the Department of Defense NDSEG fellowship.

Methods

Fabrication of PbTe & Ge particles:

PbTe & Ge subwavelength spheres of various sizes were fabricated by femtosecond laser ablation. In these experiments, we used a commercial femtosecond laser system (Spitfire, Spectra Physics) delivering ~ 1 mJ pulses with ~120 fs duration with central wavelength of 800 nm and variable repetition rate. Pulse energies ranging between 20 and 200 µJ at 20 Hz repetition rate were used in ablation experiments. More details on ablation experiments can be found in a previous work[11].

PbTe cubes were prepared per a previously reported procedure[38]. 10 mL of an aqueous 4 M NaOH (99.99%, Sigma-Aldrich) solution was prepared from pellets and refluxed at 100 °C. 221 mg $Na_2TeO_3$ (1 mmol, 99%, Sigma-Aldrich), 100 mg $NaBH_4$ (2.6 mmol, 99.99%, Sigma-Aldrich), and 330 mg $Pb(NO_3)_2$ (1 mmol, 99.9%, Sigma-Aldrich) were added quickly in turn to the NaOH solution, where upon addition of $Pb(NO_3)_2$, a black residue (PbTe) quickly formed. This mixture was then refluxed for a further 5 minutes, separated into 2 equally dispersed 5 mL aliquots, and deposited into 2 Parr autoclaves with 23 cm³ Teflon liners. The final concentration of the solution had a large impact on the final morphology of the cubes (hopper-cubes or otherwise), as described prior[38]. Into each Teflon liner, 30.5 mg cetrimonium bromide (0.083 mmol, 95%, Sigma-Aldrich) was added to the solutions which were stirred briefly. The amount of cetrimonium bromide added was found to not impact the morphology of the cubes, only that it was present. These autoclaves were sealed, heated in an oven at a rate of 1 °C/minute to 160 °C, and held for 2880 minutes. The autoclaves were cooled within the oven with no programmed cooling time. The resulting solutions were filtered to obtain grey PbTe powders (confirmed from PXRD supplementary Figure S2) which were washed with water and ethanol, and vacuum dried overnight. A 1 mg amount was taken from each, and sonicated in isopropyl alcohol for 4 hours to disperse the cubes, and drop cast onto either silicon or gold.

Single particle spectroscopy at various temperatures was conducted using an FTIR (Vertex 70, Bruker) coupled to an infrared microscope (Hyperion 3000, Bruker) using a thermal stage (THMS600, Linkam). More details on single particle spectroscopy were reported elsewhere[11]. Finite difference time domain (FDTD) calculations were performed using the Lumerical Solutions FDTD Solver, Version 8.7.3. A non-uniformal conformal mesh was used. The simulation region consisted of a semiconductor resonator over a substrate where for the case of spheres a 50-100 nm overlap with the substrate was simulated. A mesh size at least 10x smaller than the minimum wavelength in the material was used with boundary conditions of perfectly matched layers. Simulations were run for 1000 fs to ensure fields from high quality factor resonances decayed sufficiently, with boundary conditions of perfectly matched layers. The scattering cross section was computed using a total-field-scattered-field excitation source and the cross section analysis object. The reflectivity was computed by integrating the transmitted power over a monitor with an area corresponding to the numerical aperture of the experimental microscope, located above the excitation region.

Author contribution:
T.L. and J.A.S conceived the idea and experiments, analyzed the data, and wrote the paper. T.L. fabricated the semiconductor spheres, conducted the optical measurements and the Mie theory calculations. H.A.E fabricated the PbTe nanocubes. N.A.B. performed the FDTD simulations.

# Supplementary Information

## 1. Thermo-optic coefficient

The thermo-optic (TO) coefficient in the transparent regime can be defined as[1]:

$$2n\frac{dn}{dT} = (n_\infty^2-1)\left(-3\alpha R - \frac{1}{E_{eg}}\frac{dE_{eg}}{dT}R^2\right) \qquad (1)$$

where n, $n_\infty$ and T are the refractive index, the high frequency refractive index and temperature respectively, α is the linear thermal expansion coefficient, and $R = \frac{\lambda^2}{\lambda^2-\lambda_{ig}^2}$ where $\lambda_{ig}$ is the wavelength corresponding to the temperature-invariant isentropic bandgap[1], and $E_{eg}$ is the temperature-dependent excitonic bandgap.

The importance of the excitonic bandgap to the TO coefficient is clearly seen in Eq. 1. The temperature variation of the excitonic bandgap $E_{eg}(T)$ is typically the dominant contribution to the TO coefficient. For simplicity, Eq. 2 can be rewritten as:

$$2n\frac{dn}{dT} = GR + HR^2 \qquad (2)$$

Where $G = -3\alpha(n_\infty^2-1)$ and $H = -\frac{1}{E_{eg}}\frac{dE_{eg}}{dT}(n_\infty^2-1)^2$. Experimental data of many optical glass and semiconductor materials have been analyzed using the abovementioned model and the TO coefficient was found to be in very good agreement. It is clear that in the normal spectral regime (λ>$\lambda_{ig}$ and therefore R>0) the sign and value of the TO coefficient is determined by values of G and H. The contribution from G is usually negative because the thermal expansion coefficient α is positive for most materials, which dictates that G is negative. Also, the contribution of G is usually smaller than H since α is small (~$10^{-6}$/°C). Typically, the temperature variation of the excitonic bandgap $E_{eg}$ is large (~$10^{-4}$eV/°C) and negative (similarly to the bandgap energy $E_g$ which is also negative for most materials), that is as temperature increases the bandgap decreases. Since the first factor in H is also negative (-1/ $E_{eg}$) it follows that H is usually positive and is the dominant contribution to the TO coefficient. Indeed this is the case for the vast majority of semiconductors as can be seen in Figure1 (b) of main text where Si and Ge TO coefficients are plotted. Contrarily, the lead chalcogenide family PbX (X=S, Se or Te) has an anomalous negative sign of TO coefficient, which comes from the anomalous dispersion of the bandgap energy $dE_g/dT>0$ and hence also the excitonic bandgap dispersion $dE_{eg}/dT>0$. That is, the bandgap increases when temperature increases. In contrast to most materials, the PbX family has a refractive index that decreases with temperature. Although the reason for this abnormal behavior is not fully understood, several studies explain this anomaly of $dE_g/dT>0$ by an anharmonic lattice contribution (product of thermal expansion and deformation of energy potential) and an electron interaction with both acoustic and optical phonons that altogether contributes a positive value of $dE_g/dT$[2-5]

In plotting Figure 1(a) of the main text, we used the RT values for coefficients *G* and *H* of Ge, Si and PbTe detailed in ref [2]. Specifically for Si and Ge, large number of *n(T,λ)* at different temperatures and wavelengths are available in literature which made it possible to extract and express G(T) and H(T) as a quadratic function of temperature. For the Ge resonators in this work, we used the following relations in order to model the TO coefficient and *n(T)* of Ge[1]:

$$G = 0.324 - 2.092*10^{-2}T + 5.0398*10^{-5}T^2 - 4.1434*10^{-8}T^3 \qquad (3)$$
$$H = 14.336 + 9.124*10^{-2}T - 8.635*10^{-5}T^2 + 4.1382*10^{-8}T^3 \qquad (4)$$

Similarly, a relation for Si can be obtained from the work by *Li*[6] as it provides an expression for *n(T,λ)* for a large temperature range of 20K-1600K, where in ref [2] only an expression for elevated temperatures of T>293K exists.

For PbTe however, the only TO value we found in the literature was at RT[1]. Surprisingly, in our temperature range of interest (T>77K), measurements of n(T) are only available for very limited number of temperatures. Moreover, some measurements from different sources are not consistent. Consequently we used three values of n(T), at T=80K, 130K and 300K[7,8]. The imaginary part κ of the index at RT was taken from ref [9] where they integrated data of κ from several authors.

## 2. Fabrication and composition of PbTe spheres and cubes

The composition of laser ablated PbTe spheres may differ from the target wafer substrate used for fabrication, causing a change in composition and stoichiometry. To verify the composition and stoichiometry of fabricated PbTe spheres, energy dispersive X-ray spectroscopy (EDS) measurements were performed using FEI XL-30 FEG SEM with EDAX EDS. Figure S1 presents a typical elemental analysis of PbTe spheres fabricated with laser ablation. The analysis verifies the existence of Pb and Te atoms in the proper stoichiometry (49.46% Pb, 50.54% Te). The particle elemental analysis was also compared to the target wafer (PbTe 111 wafer with 99.999% purity, purchased form MTI corporation) used for ablation experiments and no composition discrepancies were identified.

**Detailed description of PbTe cube preparation:** PbTe cubes were prepared per a previously reported procedure[9]. 10 mL of an aqueous 4 M NaOH (99.99%, Sigma-Aldrich) solution was prepared from pellets and refluxed at 100 °C. 221 mg $Na_2TeO_3$ (1 mmol, 99%, Sigma-Aldrich), 100 mg $NaBH_4$ (2.6 mmol, 99.99%, Sigma-Aldrich), and 330 mg $Pb(NO_3)_2$ (1 mmol, 99.9%, Sigma-Aldrich) were added quickly in turn to the NaOH solution, where upon addition of $Pb(NO_3)_2$, a black residue (PbTe) quickly formed. This mixture was then refluxed for a further 5 minutes, separated into 2 equally dispersed 5 mL aliquots, and deposited into 2 Parr autoclaves with 23 $cm^3$ Teflon liners. The final concentration of the solution had a large impact on the final morphology of the cubes (hopper-cubes or otherwise), as described prior[9]. Into each Teflon liner, 30.5 mg cetrimonium bromide (0.083 mmol, 95%, Sigma-Aldrich) was added to the solutions which were stirred briefly. The amount of cetrimonium bromide added was found to not impact the morphology of the cubes, only that it was present. These autoclaves were sealed, heated in an oven at a rate of 1 °C/minute to 160 °C, and held for 2880 minutes. The autoclaves were cooled within the oven with no programmed cooling time. The resulting solutions were filtered to obtain grey PbTe powders (confirmed from PXRD) which were washed with water and ethanol, and vacuum dried overnight. A 1 mg amount was taken from each, and sonicated in isopropyl alcohol for 4 hours to disperse the cubes, and drop cast onto either silicon or gold. The synthesized PbTe cubes described in the main text were examined with X-ray powder diffraction (PXRD), where their structural and material composition was verified as presented in Figure S2

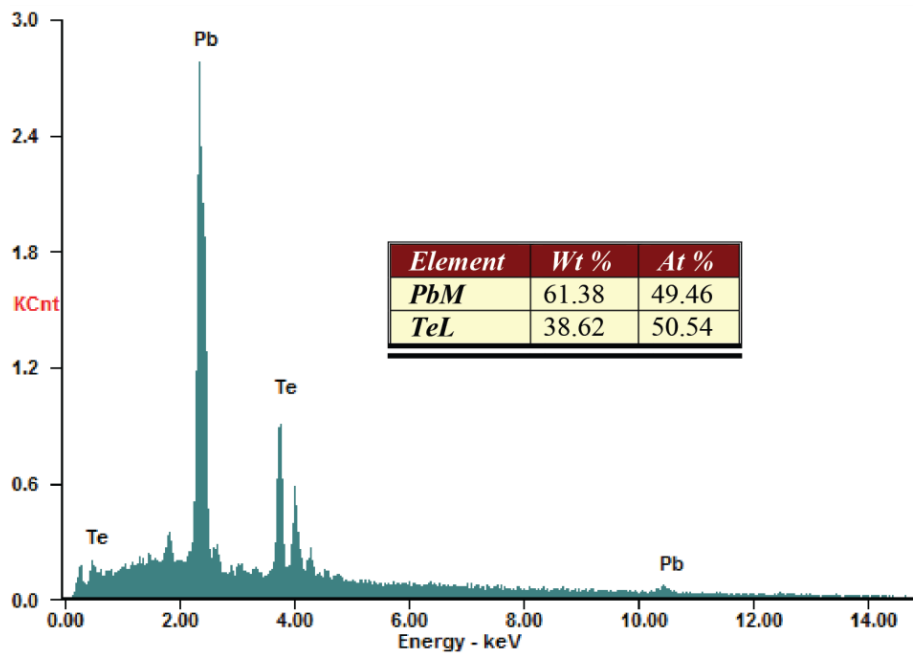

Figure S1: Energy dispersive X—ray spectroscopy elemental analysis of a typical PbTe sphere. The data was obtained for the d=1.45µm sphere described in Figures 1&2 of main text

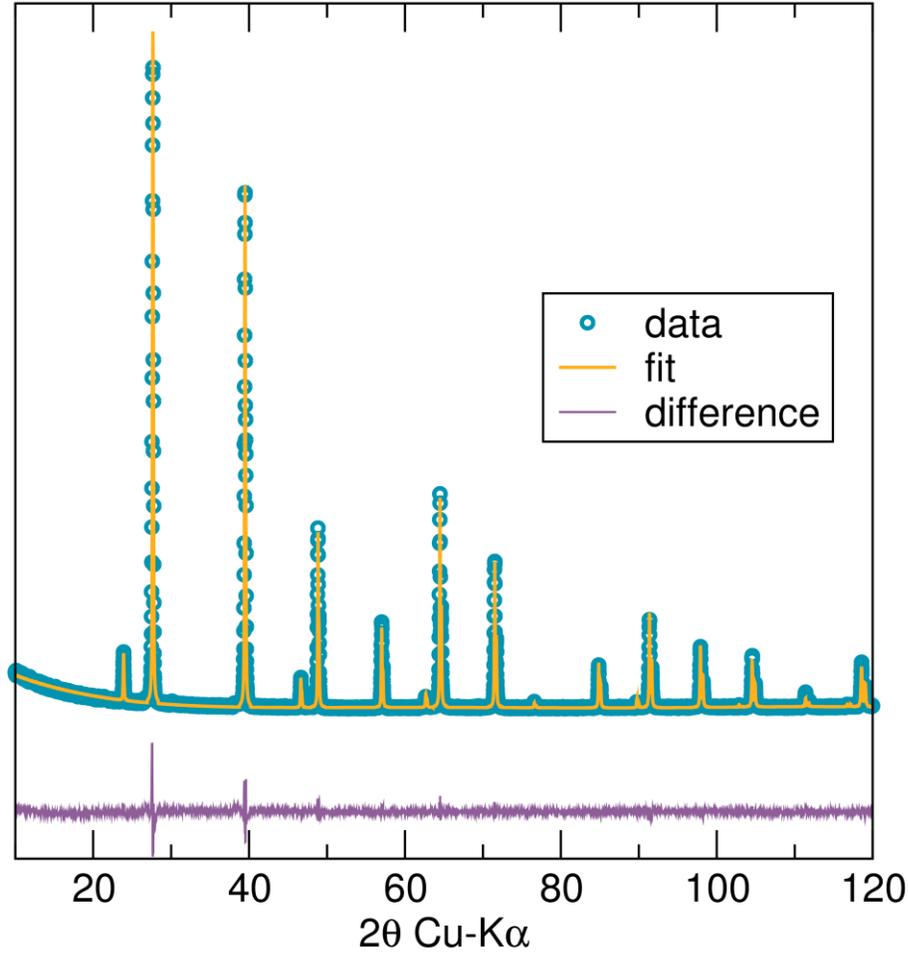

Figure S2: PXRD and Le Bail refinement of solution prepared PbTe cubes.

### 3. Resonance shifting and extracted induced index shifts

As presented in the main text, the Ge resonances shift uniformly over the measured temperature and spectral range. This is quantitatively presented in Figure S3(a) and Figure S3(c) by tracking the shift of resonance wavelengths of the first five Mie resonances with temperature compared with Mie theory predictions. The MD blue-shift and ED red-shift (and also the smaller offsets of higher order mods) in Mie theory compared to experiment, has been previously observed and explained in terms of interference with the substrate[10]. Regardless, this systematic offset does not affect refractive index differences (Δn). The slope of the measured resonance shifts with temperature and Mie theory curves match very well for both dipole and higher order modes.

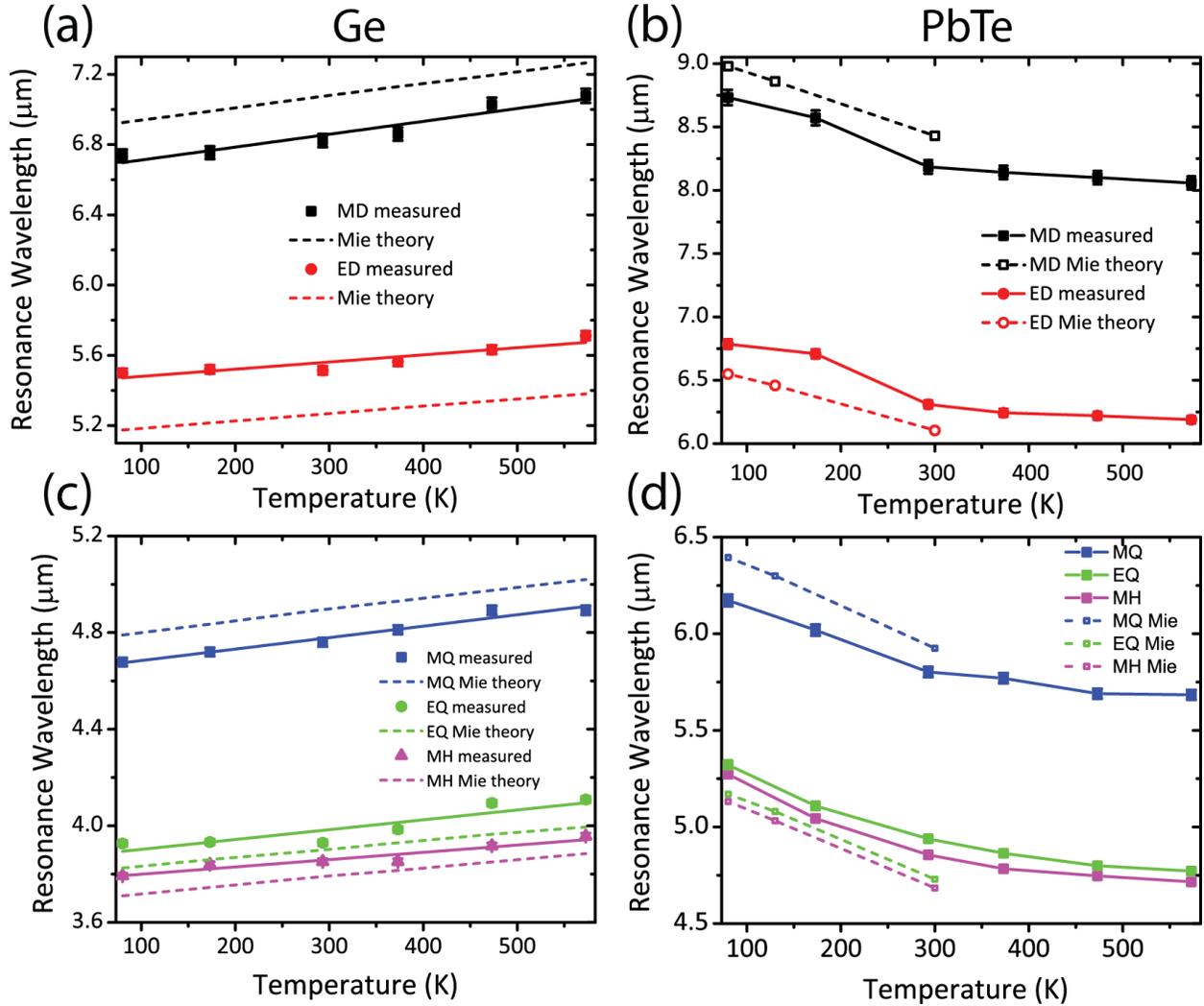

Figure S3: Multipolar resonance wavelength shift vs. temperature with Mie theory predictions in Ge sphere of diameter d=1.7µm on Ge substrate and PbTe sphere of diameter d=1.45µm on PbTe substrate. (a) & (b) Dipole modes. (c) & (d) High order modes.

The tunable PbTe resonances exhibit different characteristics as demonstrated in the main text. In Figure S3(b) and Figure S3(d) we plot the resonance TO tuning of the first five multipoles. All modes follow the same trend: the largest shifts appear between 80K and 293K followed by a moderate shift from 293K to 573K. As mentioned previously, literature reports of refractive index data of PbTe at various temperatures is limited[7,11,12] precluding calculations across the measured temperature range. We were therefore unable to plot Mie theory predictions beyond three discrete datasets at 80K, 130K and RT[7]. However, shifts of Mie resonances between 80 and RT match literature values (the slop of experiment and Mie theory curves match well between 80 and RT). The normalized tunability of the MD mode between 80K and 293K, discussed in Figure 1d of the main text, was -0.65. Interestingly, due to its narrower linewidth, the normalized tunabilty of the electric dipole (ED) mode between 80 and 293K (Normalized tunability=1.5) is more than doubled than the MD mode tunability. This normalized tunability is much larger than previous reports in single Mie or plasmonic antennas[13]

The extracted temperature induced index shift Δn taken from several multipolar resonances in several Ge spheres, is presented in Figure S4(a). The procedure for obtaining these values is as follows: The effective refractive index at a particular resonance is obtained from the resonance condition derived from Mie theory[10]: $2\pi rn/\lambda$=constant, where r is the particle radius, n is the refractive index and λ is the free space resonance wavelength. Measuring the particle radius and resonance wavelength allows to extract the refractive index. Then the induced Δn(T) for each resonance is calculated relative to the index at the lowest measured temperature (80K), that is, Δn(T)=n(T)-n(T=80K). The expected and measured induced index shift match well, which is manifested by a similar flat maximum Δn≈0.2 for the full temperature swing of 80-573K and across the entire MIR spectral range. This moderate induced Δn is sufficient to provide more than a linewidth of tunability for the high order modes in Ge but may require a large temperature swing ΔT. Also, this moderate Δn will induce a larger resonance shift for larger particles exhibiting resonances at longer wavelengths. This is due to the Drude dispersion of n(λ) which drops with~ $\lambda^2$ and although Δn is relatively constant with wavelength, the Δn/n ratio becomes larger for longer wavelengths leading to larger wavelengths shifts. This is illustrated in Figure S5 where the TO shift in a Ge sphere of diameter d=3.2μm is presented. The wavelength shift of the MD mode is 630nm across the full measured temperature difference of 80-573K leading to normalized tunability =0.48, which is larger than the normalized tunability=0.34 of the d=1.7μm resonator presented in Figure S3 (and Figures 1. & 2. of the main text) over the same temperature range. The normalized tunability improves when moving to high order modes. For instance, the normalized tunability of the MQ and EQ modes in the d=3.2μm particle of Figure S5 are 0.77 and 1.58, respectively.

The chromatic dispersion of the TO coefficient in PbTe that was apparent in Figure 1(b) of main text, also emerges in Figure S4(b). The extracted Δn follows a similar functional behavior as the calculated TO coefficient. It also shows that the largest jumps in Δn occur between 80K to RT where a more thorough examination reveals that dn/dT peaks between 80K and 173K. Induced index shifts as high as Δn≈0.7 are obtained at shorter wavelengths and these values are expected to increase for wavelengths approaching the excitonic bandgap following Eq 2. Figure S4 presents the induced Δn in PbTe cubes on both gold and on Si substrates. The TO resonance tuning characteristically behaves similar to the spherical resonators with the largest shift observed between 80K and 293K. However, the observed induced index shifts Δn in cubes are up to 15% larger than that in spherical resonators. This may be attributed to the greater degree of crystallinity of the cubes compared to the spherical resonators.

Altogether, we see a consistent behavior in both laser ablated spherical resonators and solution processed cubic resonators that shows that the TO in PbTe is significantly larger at low temperatures than at RT. We also identify the TO spectral behavior expected from theory, namely increasing values for wavelengths approaching the excitonic bandgap. These two factors contributed to the extremely large local dn/dT=0.0134 that was observed at 173K-163K at $\lambda$=4.9 µm in the spherical particle presented in Figure 3(d) of main text. In comparison, for the EQ mode in that spectrum and same temperature range 173K-163K, we obtained dn/dT=0.007 at $\lambda$=6 µm, which is five times larger than the RT value dn/dT=0.00136.

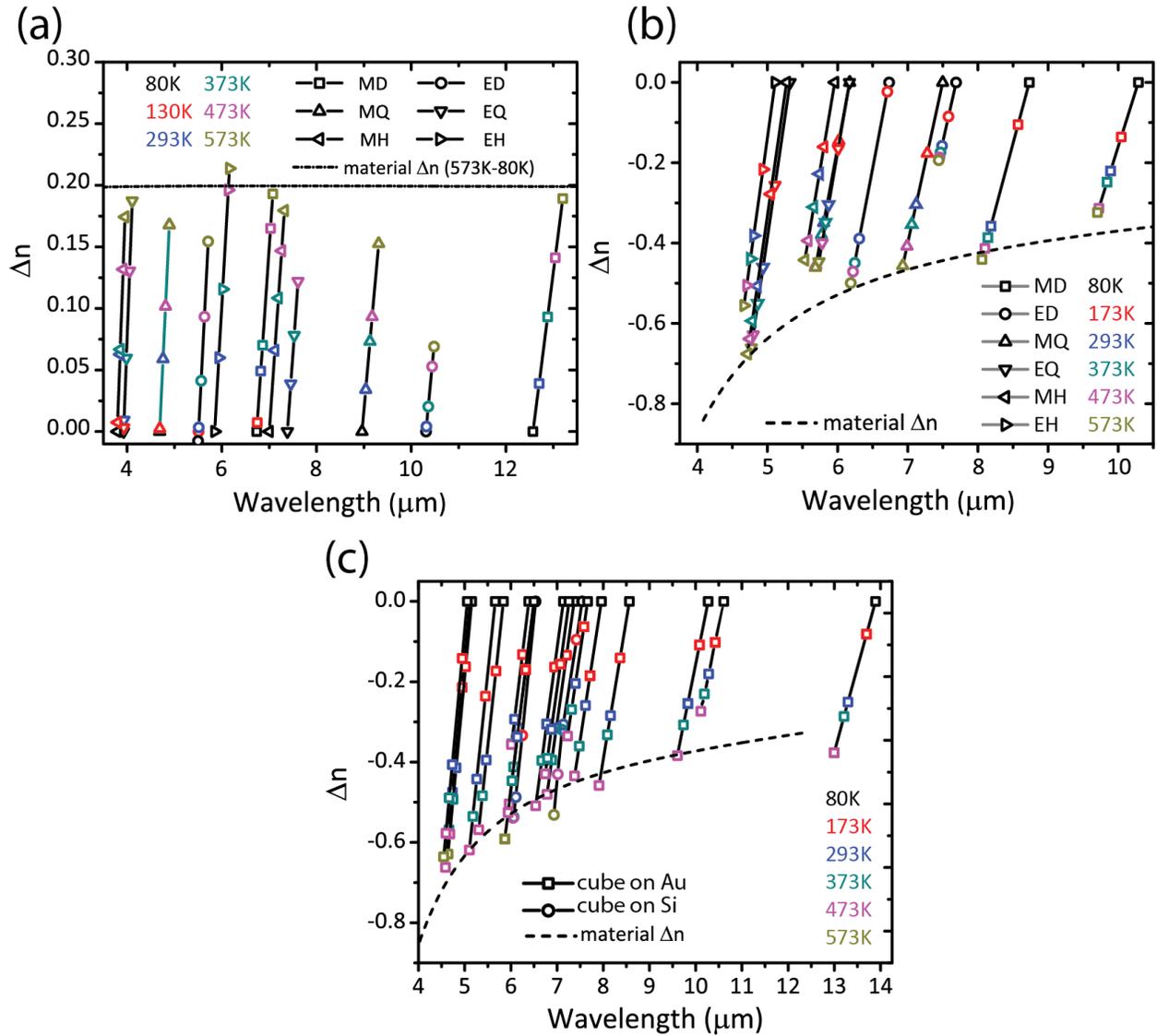

Figure S4: The extracted TO induced index shift in (a) Ge spherical resonators on Ge substrate. (b) PbTe spherical resonator on PbTe substrate. (c) PbTe cubes on Si and gold substrate. All index shifts are calculated with the respect to the index at 80K. Each color corresponds to different measured temperature. The different symbol in (a) and (b) correspond the different multipole modes (MD, ED, MQ, EQ, MH or EH) in spherical resonators. The dashed line corresponds to the functional behavior of Δn in the material.

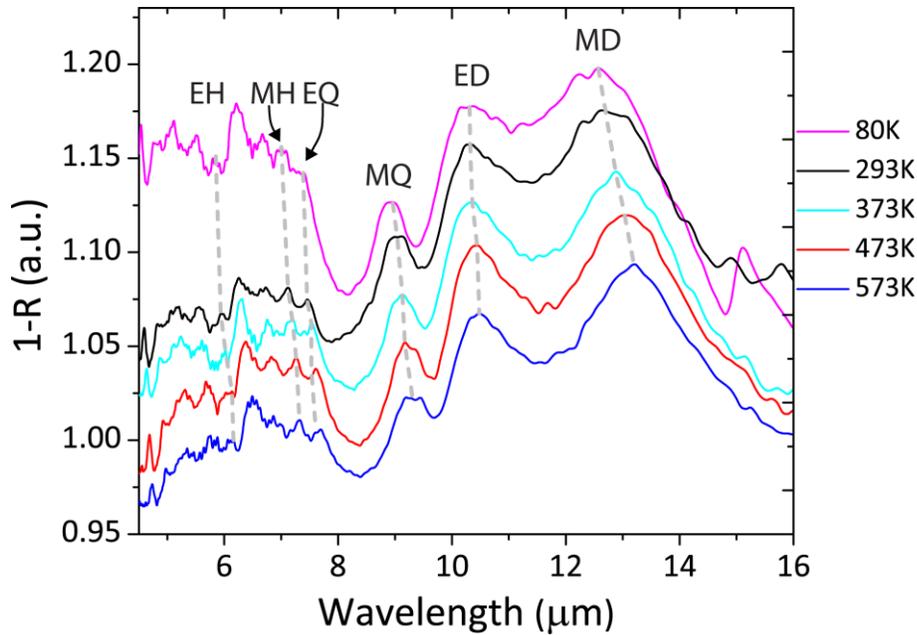

Figure S5: Spectra of a Ge sphere of diameter d=3.2µm on Ge substrate for different temperatures. The moderate induced index of Δn≈0.2 in Ge between 80 and 573K induces a resonance shift of up to 700nm in the MD mode at long wavelengths. The temperature induced index shift is also sufficient to shift the higher order modes (quadruples and hexapoles) by more than a linewidth.

## 4. Ultrahigh tuning of high-Q resonances in spherical and cubic PbTe resonators

High-index, low-loss and large TO effect are all combined in PbTe to allow ultrahigh tunability of resonances. Figure S6 presents the measured infrared spectrum of the d=1.76µm sphere (discussed in Figure 3 of main text) along with full Mie scattering cross-section (solid red) and the Mie scattering cross-section when only ED, EH and magnetic octapole (MO) modes only (dashed red). Very good match exists between the theoretical curves and experiment. Specifically, the location of the sharp resonance feature (shaded grey area) is in excellent match between the curves. Mode analysis obtained by Mie theory shows that this resonance feature arises due to the mode coupling of the ED and EH modes, with a smaller contribution from the MO modes. This unique high-Q feature of in PbTe spheres is the basis of ultrahigh tunability demonstrations in spherical resonators.

The temperature dependent spectra of single cubic PbTe resonator of side a=1µm on a Si substrate (presented in Figure 3(a) of main text) is presented in Figure S7. The shifting of resonances follow similar behavior of spherical resonators exhibiting a peak TO tunability between 80K and RT.

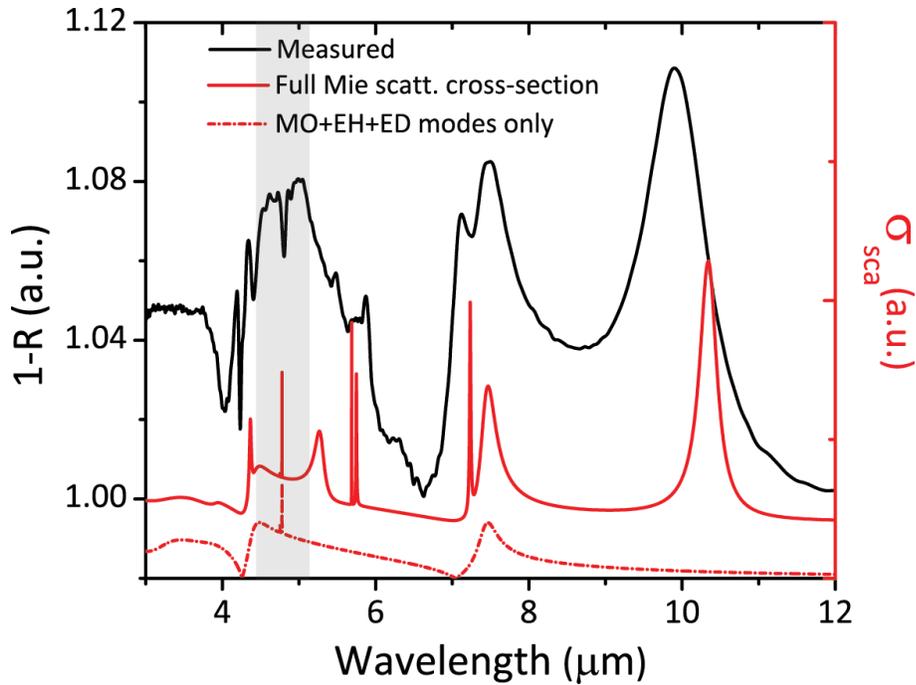

Figure S6: Measured infrared spectrum of the d=1.76µm sphere discussed in Figure 3 of main text along with full Mie scattering cross-section (solid red) and the Mie scattering cross-section coming of the combined ED, EH and magnetic octapole (MO) modes only (dashed red). The theoretical curves were shifted and rescaled for visibility. Very good match exists between the theoretical curves and experiment. Specifically, the location of the sharp resonance feature is in excellent match between the curves. The mode analysis obtained by Mie theory shows that this resonance feature arises due to the mode coupling of the ED and EH modes, with a smaller contribution from the MO modes.

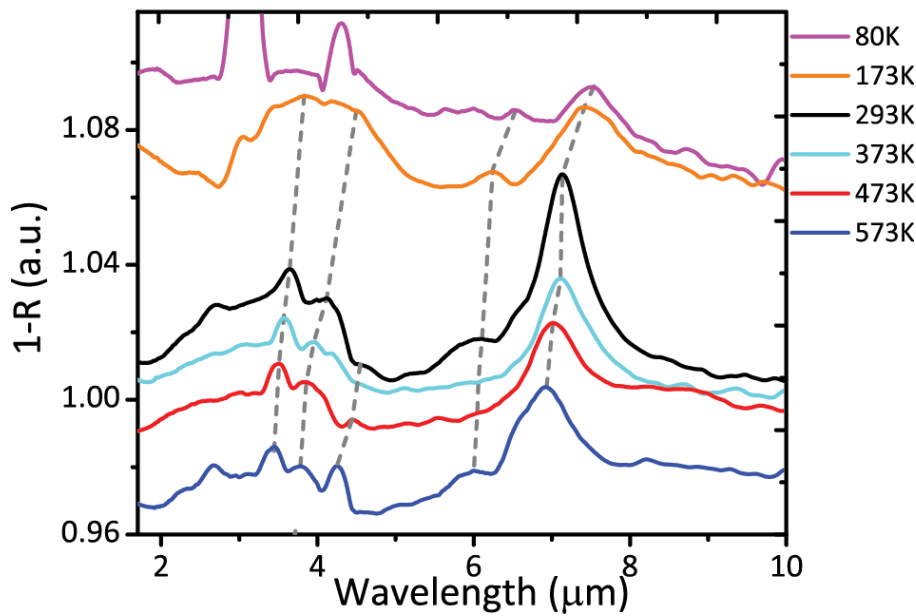

Figure S7: TO tuning of single cubic PbTe resonator of side a=1µm on a Si substrate

As we show in the main text, cubic resonators on gold exhibit high-Q resonance features. Figure S8(a) and S8(b) present the temperature dependent infrared spectrum of PbTe cubes of sides a=1.76µm and a=1.35µm, respectively, on gold substrate. Similar to the high-Q features in the spherical resonators, these features can be tuned by several linewidths with a peak TO coefficient observed between 80-293 K. This is further illustrated in Figure S8 (c) and S8 (d), where a zoom-in on the high order modes of these resonators is presented, for small temperature variations around T=173K and below. These resonances are tuned by more than a linewidth (normalized tunability>1) for temperature swings as low as ΔT=30.

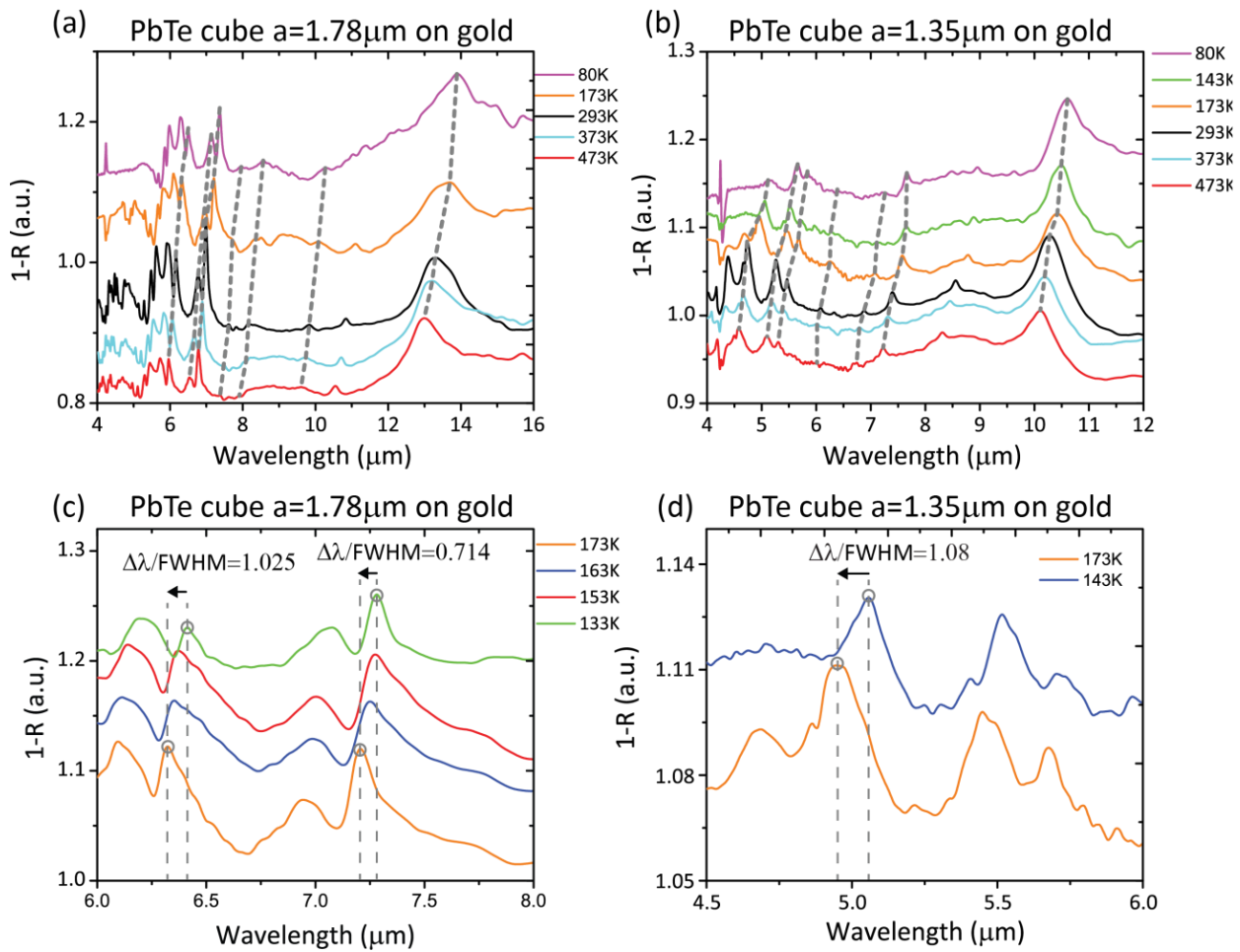

Figure S8: TO tuning of high order high-Q resonances in cubic PbTe resonators on gold (a) Temperature dependent infrared spectrum of a PbTe cube of side a=1.78µm on gold substrate. The dashed lines are a guide to the eye. (b) Temperature dependent infrared spectrum of a PbTe cube of side a=1.35µm on gold substrate. The dashed lines are a guide to the eye. (c) & (d) zoom-in on linewidth tunability of high order modes with small temperature span of the cubes in (a) & (b). The normalized tunability Δλ/FWHM is labeled in each panel. The resonance quality factors shown in (c) and (d) are Q~60 and Q~50, respectively.

## 5. Ultra-sensitive switching in cubic metasurface arrays on PEC

Figure 5(c) of main text, presents the infrared spectrum of PbTe cubic metasurface array on PEC. As we demonstrated in Figure 5(b), ultra-wide tuning can be achieved with these resonances. In Figure S9, we present tuning by multiple linewidths with the cubic metasurface spectra detailed in Figure 5(c) of main text. The inset shows how the blue spectrum can be tuned by more than a linewidth, with steps of induced index of Δn=0.005. This Δn=0.005 is equivalent to temperature swing of ΔT=0.37K at T=173K or ΔT=3.33K around RT.

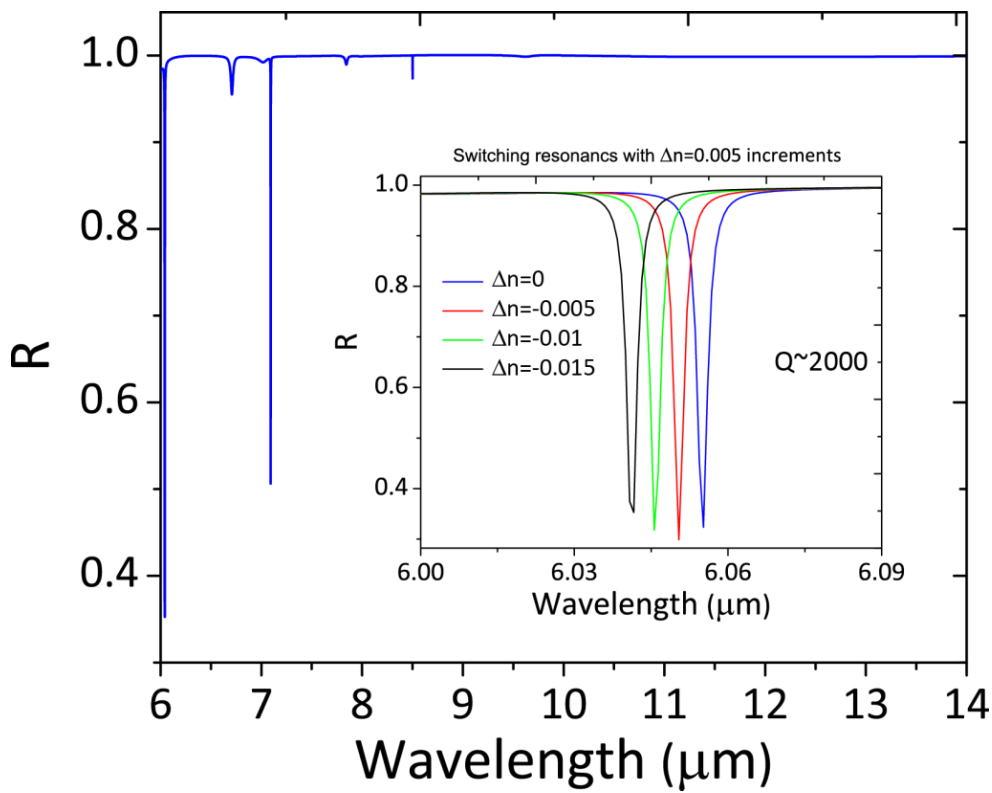

Figure S9 : Ultra-wide tuning in cubic PbTe metasurfaces on PEC. The inset shows how the blue spectrum can be switched, i.e., tuned by more than a linewidth, with induced indices of Δn=0.005. This Δn=0.005 is equivalent to temperature swing of ΔT=0.37K.